# Towards Automatic Personality Prediction Using Facebook Like Categories

Raad Bin Tareaf, Philipp Berger, Patrick Hennig and Christoph Meinel

Hasso-Plattner-Institute

University of Potsdam, Germany

firstname.lastname@hpi.uni-potsdam.de

*Completed Research Paper*

*We demonstrate that effortlessly accessible digital records of behavior such as Facebook Likes can be obtained and utilized to automatically distinguish a wide range of highly delicate personal traits including: life satisfaction, cultural ethnicity, political views, age, gender and personality traits. The analysis presented based on a dataset of over 738,000 users who conferred their Facebook Likes, social network activities, egocentric network, demographic characteristics, and the results of various psychometric tests for our extended personality analysis. The proposed model uses unique mapping technique between each Facebook Like object to the corresponding Facebook page category/sub-category object, which is then evaluated as features for a set of machine learning algorithms to predict individual psycho-demographic profiles from Likes. The model properly distinguishes between a religious and non-religious individual in 83% of circumstances, Asian and European in 87% of situations, and between emotional stable and emotion unstable in 81% of situations. We provide exemplars of correlations between attributes and Likes and present suggestions for future directions.*

**Keywords:** Personality Prediction, Machine Learning, Social Networking Sites, myPersonality Dataset.

## INTRODUCTION

Personality is determined as a set of characteristics which make an individual unique, and the study of personality considered as a central aim of psychology [1]. One of the most influential and generally accepted personality theories is the big-five personality theory, which envelope five basic traits: Extraversion (sociable vs shy), Agreeableness (friendly vs uncooperative), Conscientiousness (organized vs careless), Openness (insightful vs unimaginative), and Neuroticism (neurotic vs calm) to compose human personality [2]. With the wide spread of social networks sites nowadays, Facebook becomes one of the most popular social networking services in the world. More than 1.3 billion users are daily active as on average of June 2017[1]. As a consequence, Facebook plays a big role in people' normal life. Thus, the platform provides an ideal online platform for personality research and relative application [4].

The information revealed through a personality assessment can be used in numerous applications. These include but not limited to, advertisements alignment, marketing campaigns adjustment, and supporting bloggers in narrowing down their target audiences based on community pre-detected personality traits [3] [5]. There are also many other applications that can take an advantage of personality recognition systems. For instance, a company selling guns can selectively show an advertisement describing their products as a sign of strength and force to extroverted people, while showing statistics of burglary and highlighting the apparent safety improvements to neurotic and anxious people.

In recent years, the interest of the scientific community in personality recognition has grown very fast [6] [7]. Meanwhile, predicting user's personality through social networks is not an easy task. One of the critical factors that affect personality detection at the scale of social platforms is the predictive accuracy as an outcome of limited available training data. One of Facebook advantages (easy access to large amounts of personal data) introduce serious ethical challenges that have yet to be addressed in a pragmatic manner by the applicable legal and ethical guidelines.

Several authors have looked at the Big Five personality traits of Facebook users [6] [7] [8] [9]. However, there is fewer work that analyze the usage of like button in the Facebook social platform in term of personality prediction. Likes can be used by Facebook users to endorse content such as status updates, comments, photos, links shared by friends, Facebook pages, or external Web sites. Endorsements also result in users receiving updates on a given piece of content, such as comments on a liked status update. Likes were introduced by Facebook in February 2009. Likes can be used to categories users across a large variety of personal characteristics, political affluence, sexual orientation, or cultural ethnicity [10].

---

[1] https://newsroom.fb.com/company-info/

Results indicate that extra diverse information about user's personality can be determined computationally by extending the data with metadata such as user browsing history.

Studies that perform Natural Language Processing to understand users' language on social platforms leads to predict a user's personality traits precisely [11]. Others analyzed semantical features rather than syntactical features to judge human's personality [12]. It shows that not only the number of posts and statistical information of users can be used to assess personality traits. In contrast, the way in which users phrase their posts entails a lot of information about a user's personality. In fact, they were able to improve existing computer-based personality assessments by supplementing the syntactical features with their semantical ones. However, the metadata of the like object itself on the Facebook social platform has not been studied in any research yet.

The structure of this paper is as follow: The previous INTRODUCTION Section presents the related works of predicting users' personality within various social networking sites. Section METHODOLOGY provides an overview of the followed and applied research methodology. In Section IMPLEMENTATION, we demonstrate the datasets we used and gives a solid view of the actual implementation. Where in Section RESULTS, we illustrate the performance insights for each algorithm we applied. Finally, In Section DISCUSSION we summarize final results with redirection of future work for the task of users' personality prediction.

## METHODOLOGY

The purpose of this study is to present a model that predicts user's personality scores by analyzing their online social fingerprints and developed a web-based application to access this prediction. We aim to utilize the hierarchy that Facebook employs to categories pages as features, then we used this features to train our machine learning models to predict user's Big Five scores for each of the personality traits. We observed a hidden relationship between the metadata of the Facebook like's objects. Therefore, we decided to investigate this relationship and map each like object to it corresponding category using

Facebook Graph API2 and evaluated different classifiers (boosted trees, linear regression, k-nearest neighbor, and neural networks) for predicting personality traits of each like category.

**IMPLEMENTATION**

As a foundation for our research we used the "big5" and "user likes" datasets from the "myPersonality project" [10]. This dataset contains information about Big Five personality scores and Facebook likes for more than 700,000 Facebook users. The personality scores are represented on a scale according to the Big Five personality traits "openness to experience", "conscientiousness", "extraversion", "agreeableness", and "neuroticism". It also includes information about the size of the questionnaire that users responded to associated with their ids.

This dataset can be joined with the data provided in the "user likes" dataset, which is a mapping from user id to Facebook like id. Since the ids are the original Facebook page ids, the missing information can be queried from the Facebook API as presented in figure 1. In order to crawl the needed mapping data, we developed a query bot that used the Facebook Graph API to get all available information about the likes, such as, category, subcategories, engagements, verification status, price ranges and more. However, for this study, we focused on the category and subcategories of the like objects. We then used the queried information to create a new dataset which maps a user id to the corresponding Big Five scores and the number of likes that user has in each category and subcategory. This allows us to study the relationship among these metadata of like objects and present novel method in predicting users' personality from Facebook platform.

---

[2] https://developers.facebook.com/docs/graph-api

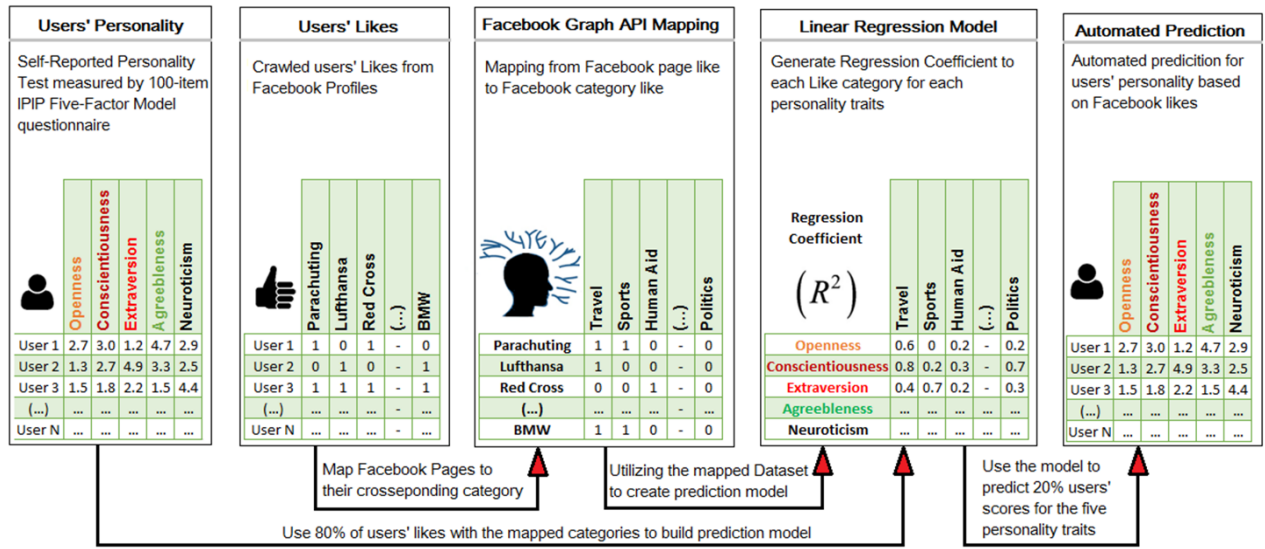

Figure 1. Proposed Model for Predicting the Big 5 Personality Traits.

We used two different sampling methods for two different objectives. First, we used random sampling which gives each element in the dataset an equal chance of being selected for the test set. This represents the best possible distribution but can lead to minorities of not being represented equally in the test set. Second, we used stratified sampling, meaning that the population is partitioned into non-overlapping groups, for example, we used buckets for each Big 5 category containing similar user objects as presented in Figure 2. Samples are then randomly picked from these groups, leading to an equal distribution from each partition.

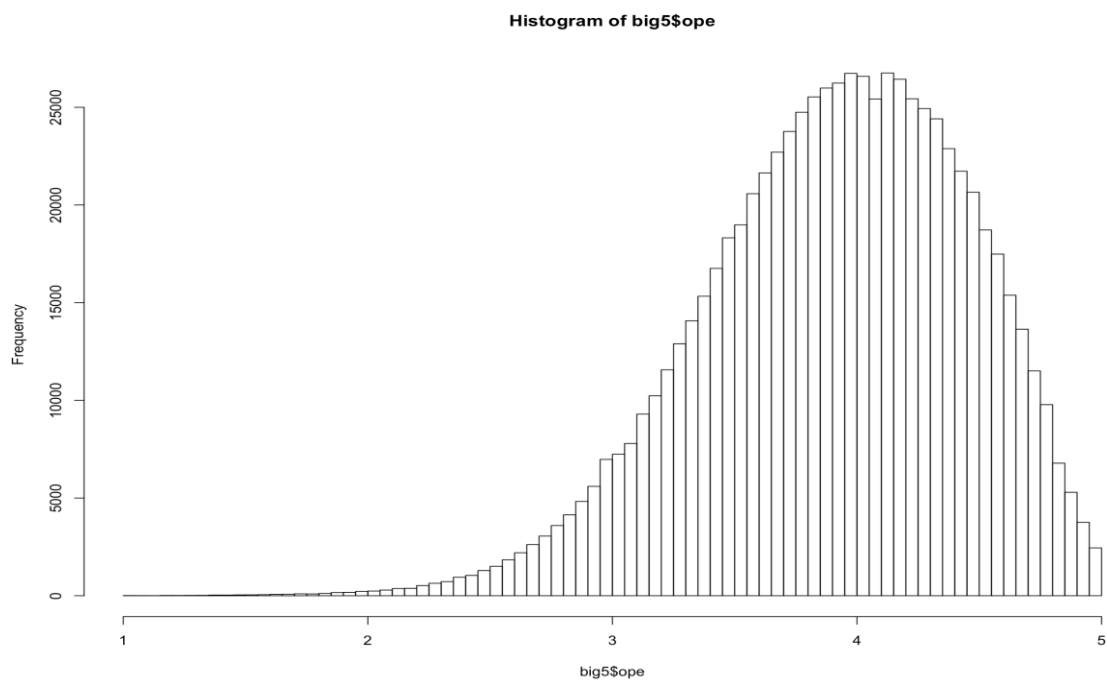

Figure 2. A histogram that reflects the Openness Distribution in the Dataset.

Since the classification and regression algorithms we used to make predictions based on a quantitative scale, huge amount of elements are treated differently than smaller amount. However, when a user has 30 likes in politics and 300 likes in sports, he might be less interested in politics than another user who has solely 25 likes in politics. Therefore we turn the absolute scale into a relative scale. A value of 0.1 in politics means that a 10% of his likes are of the politics category. Normalising the features improved our RMSE values by about 5% as we present in RESULTS section.

We explored several machine algorithms to find the optimal classifier for our task. Therefore, we experimented Random Forest algorithm as a classification algorithm based on decision trees. To create the decision tree, we mapped the data into an n-dimensional space where each dimension stands for one like category. The algorithm then tries to find a decision boundary and divides the dataset into two non-overlapping partitions. This continued until a remaining group can be perfectly separated which could result in very specific category buckets.

In a regression algorithms setting, we predict the Big 5 personality traits on $1 - 5$ linear scale using different learning algorithms. Linear regression is a learning technique that model relationship between a numeric variable with one or more feature variables. In our case, the features are the number of likes a user has in a certain category. Assuming we only have the feature ''numLikesBoxingStudio''and we want to predict the openness of a user. We therefore, try to find values for $\Theta_0, 1$ so that the following function is optimal:

$$\Theta_0 + \Theta_1 * numLikesBoxingStudio = opennessScore$$

We constructed an explanatory example and visualized it in Figure 3. These relationships are found by linear regression and modelled through the predicting function.

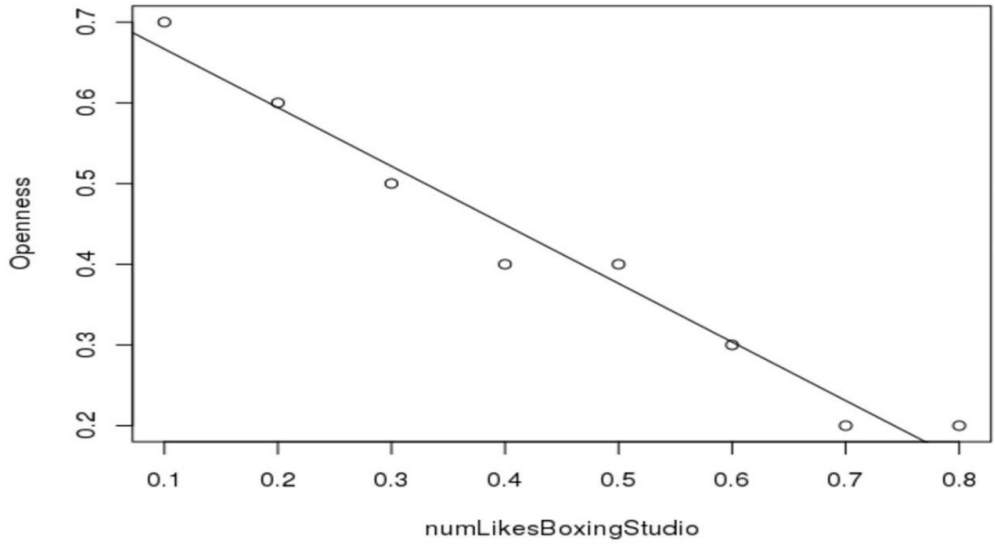

Figure 3. A Linear Regression representation for the relationship between feature and openness personality trait.

# RESULTS

To compare and evaluate prediction performance of different algorithms and models, we calculated various metrics that are specific to the class of algorithms. Classification algorithms are evaluated differently than regression algorithms because they serve complete different tasks. Therefore, we evaluated the Multi-Class classification algorithm with precision and recall and evaluated the regression algorithm with Root Mean Squared Error (RMSE). The Mean Squared Error (MSE) and Root Mean Squared Error (RMSE) both are metrics to measure the difference between predicted value and the actual value. The closer the value is to zero, the better is the model in predicting the target value. The RMSE describes by how much the predicted value on average deviates from the actual value of the observation (Equations (1) and (2)).

$$(1) \ MSE = \frac{1}{n} \sum_{i=1}^{n}(y^i - yi)^2$$

$$(2) \ RMSE = \sqrt{MSE}$$

To decide which group of algorithms fits our task, we adopted to solve the problems by training two models: "Random Forest", typically used for classification tasks, and "Linear Regression", which is used to predict continuous values. On the early stages, regression models outperformed classification models. We were able to achieve precision values of about 40% with the optimised set of features for

"Random Forests", but the task at hand is not a classification task. Additionally, the class labels do not provide a sufficient information about the person himself. Therefore, we decided to focus solely on algorithms that predict continuous values.

There are a variety of models that can be used to predict continuous values. In our research, we evaluated four different algorithms: Regression trees, linear regression, k-nearest neighbors, and neural networks. We used the RMSE function to assess which algorithm outperforms the others in predicting personality traits. Figure 4 presents the results of our experiment based on likes mapping feature.

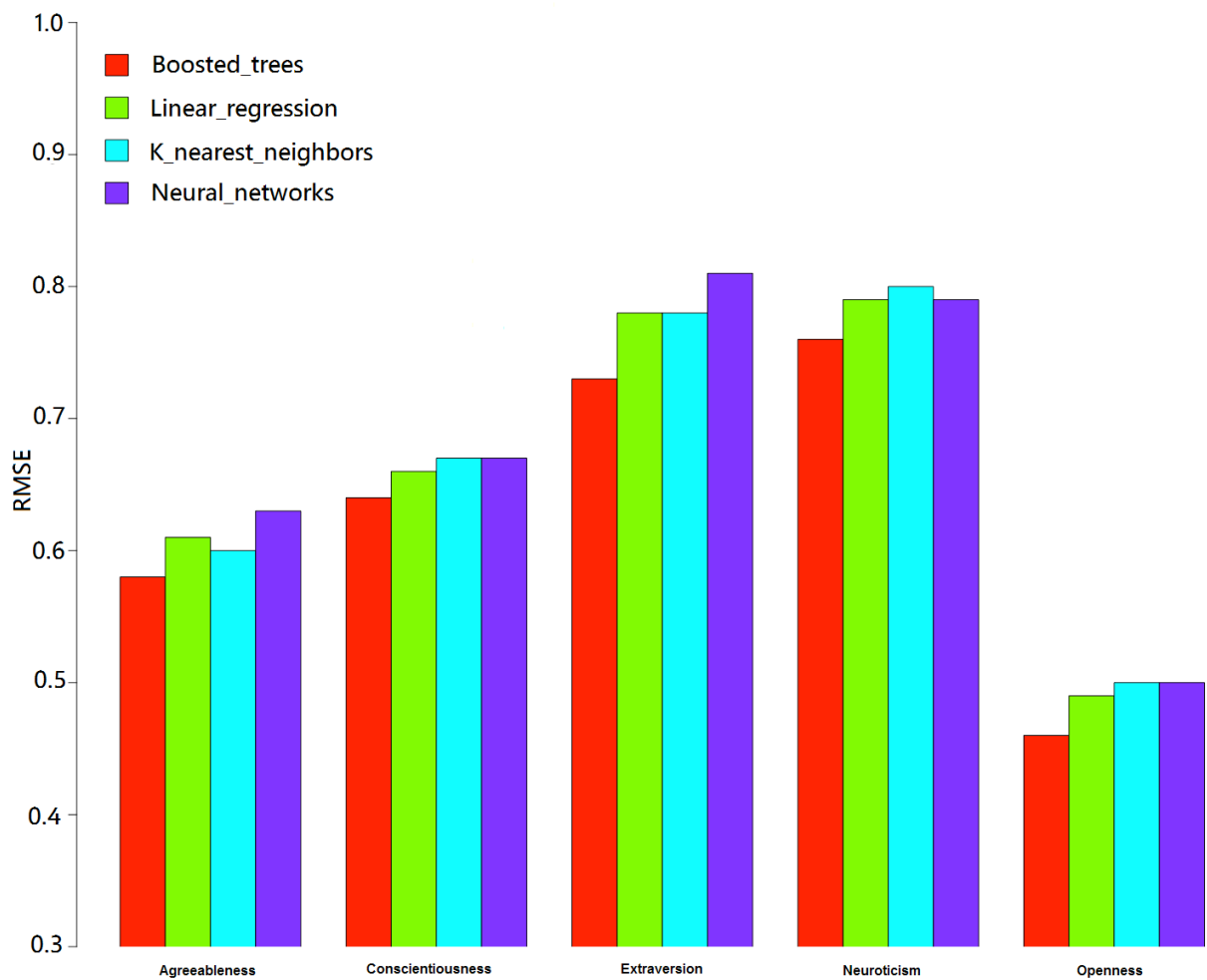

Figure 4. Root Mean Squared Error (RMSE) function for (boosted trees: Red, linear regression: Green, K-nearest neighbours: Blue, and neural network: Purple) algorithms in predicting each of the Big 5 personality traits.

The extensive experiments revealed the best-performing algorithm among the four which is the boosted trees, followed by linear regression. While these algorithms offer a lot of flexibility in the tasks they can perform, they usually introduce unnecessary complexity. In our case, we deal with a rather traditional regression task. The linear regression models are quite capable of high-quality predictions, and in fact performs even better with the feature set we optimized it for. Since the output we look for has to be limited in a 1 – 5 scale, we have to set up the logical boundaries of the setting. We, therefore, clipped the output to have all values above five to be rounded down to five and all values below one to be rounded up to one.

The algorithm we used for boosted trees is called "xg- boost"[3]. It has gained popularity in recent years by being the winner algorithm in many of machine learning competitions. Boosted trees aim to automatically address the weaknesses of the model during the training stage. In a given step, it computes which training segment struggle in prediction and then generates a tree specifically trained to better predict those observations. It gradually computes a fairly sophisticated model that consists of multiple smaller trees each optimised to predict certain characteristics of the training set.

The k-nearest neighbor model we built was able to compete with linear regression, but could not quite achieve the same exact metrics. We experimented with different numbers for K, and found $10 <= k <= 15$ yields the best results. Based on various experiments with different forms of penalties we decided to consider the number of non-overlapping categories in which the users have likes and use this as the basis for the weighted penalty. Consequently, the more categories exist in which the observation has likes, but the neighbor does not, the higher is the penalty. One way to further improve the model is to analyse the importance of each feature and significantly reduce the number of features to the ones that are most impactful on prediction performance. Currently, users might be penalized heavily for having similar likes which are assigned by Facebook to different subcategories. Generalising these might help improve prediction performance because distance-based metrics would assign them closer together. The fourth algorithm we trained models for is a neural network. A neural network can be modelled as a network of perceptron's in which the first layer gets the features as input and the succeeding inner layers get the output of the previous perceptron's as input.

---

[3] https://github.com/dmlc/xgboost

To validate our results, we divided the dataset into a test set containing 20% of the data and a training set containing the rest. Results show that we are able to predict a user's Big Five "openness" score within about 8% on average with the boosted trees algorithm. The "openness" personality trait can be predicted with the smallest error, indicating that it correlates mostly with pages that a user liked on Facebook. To make the result of our research accessible to public users; we set up a web-based prototype on which users can predict their own personality using their Facebook likes data.

## DISCUSSION

We have presented a novel approach in revealing the Big 5 personality traits of a user by predicting them from the actual Facebook pages like's list. According to the results derived from RMSE evaluation function, the proposed model can predict the "openness" personality trait with the smallest error, indicating that openness personality trait particularly correlates with pages that a user liked on Facebook. This allows for a quite accurate assessment of a person's personality traits and can be used in a wide variety of fields, for example, in political or marketing campaigns.

We examined which parameters impact the results and we observed that total number of likes a user holds play a dominant role in our experiment. Since we use each like of a user and create features out of them, the more likes a user has, the better the prediction is. On the other hand, by filtering our dataset for a minimum number of likes, we significantly reduce the number of observations that we can work with. In fact, the dataset can be reduced by about 75% when specifying a minimum number of 250 likes. Consequently, a much smaller part of the data is available to train the models, which in turn negatively influences prediction performance as presented in Figure 5.

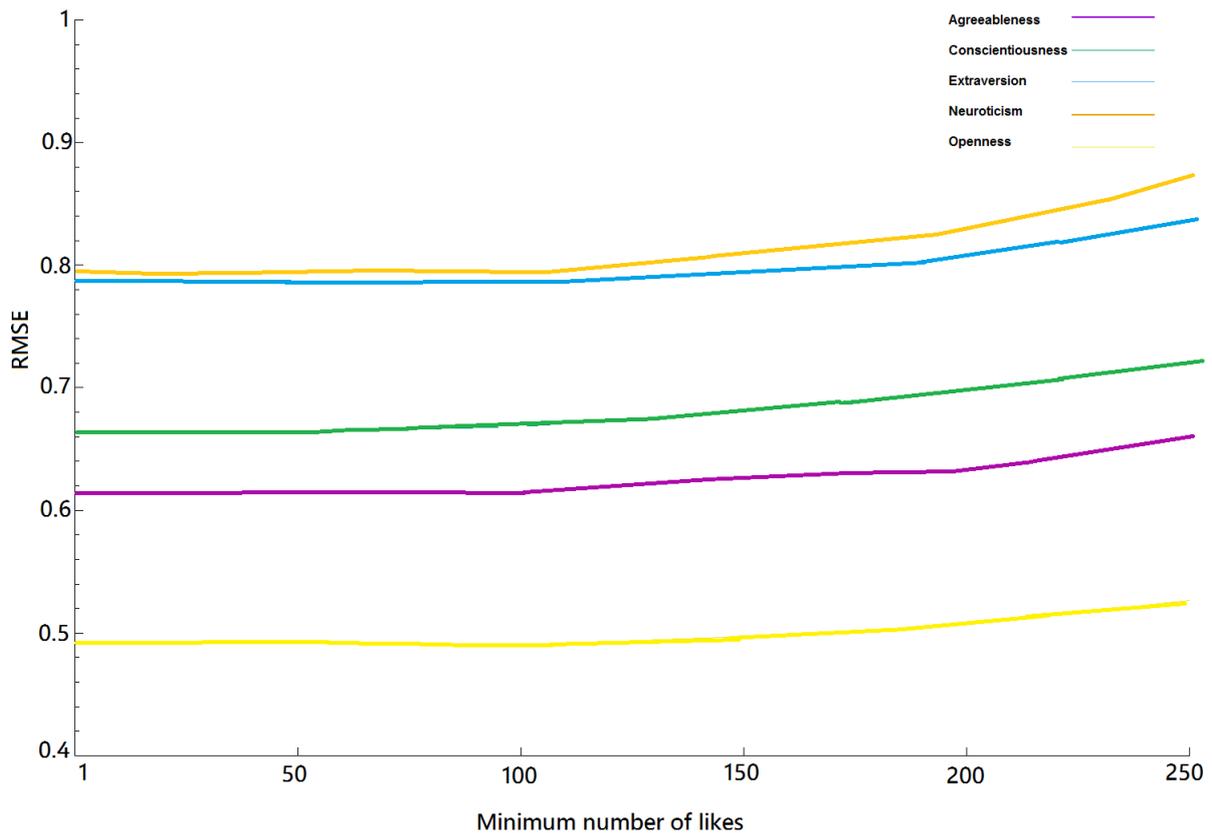

**Figure 5. Root Mean Squared Error (RMSE) value for varying minimum number of likes with maximum size of training set.**

The prior hypothesis is verified by the second round of experiments. In this round, we used a fixed-size training set in every round. Therefore, in contrast to the previous experiments, we train the models on the same number of observations, no matter how many observations exist in the whole dataset with the respective number of likes. As presented in Figure 6, prediction performance, in that case, does improve when filtering only for users with a higher number of likes.

Since the information revealed within the metadata of the like objects is only a small part of the total information a user leaves on social networks, the learning models we described can now be integrated into a learning ensemble that considers other relevant information as mentioned in the introduction section to better build personality detection systems. Last but not least, we are currently extending and magnifying the metadata information about liked pages. One idea is to enrich the data by looking up the liked entities in databases like DBpedia[4]. In cooperation with domain experts it can be used to develop insights, and consequently, generate augmented features.

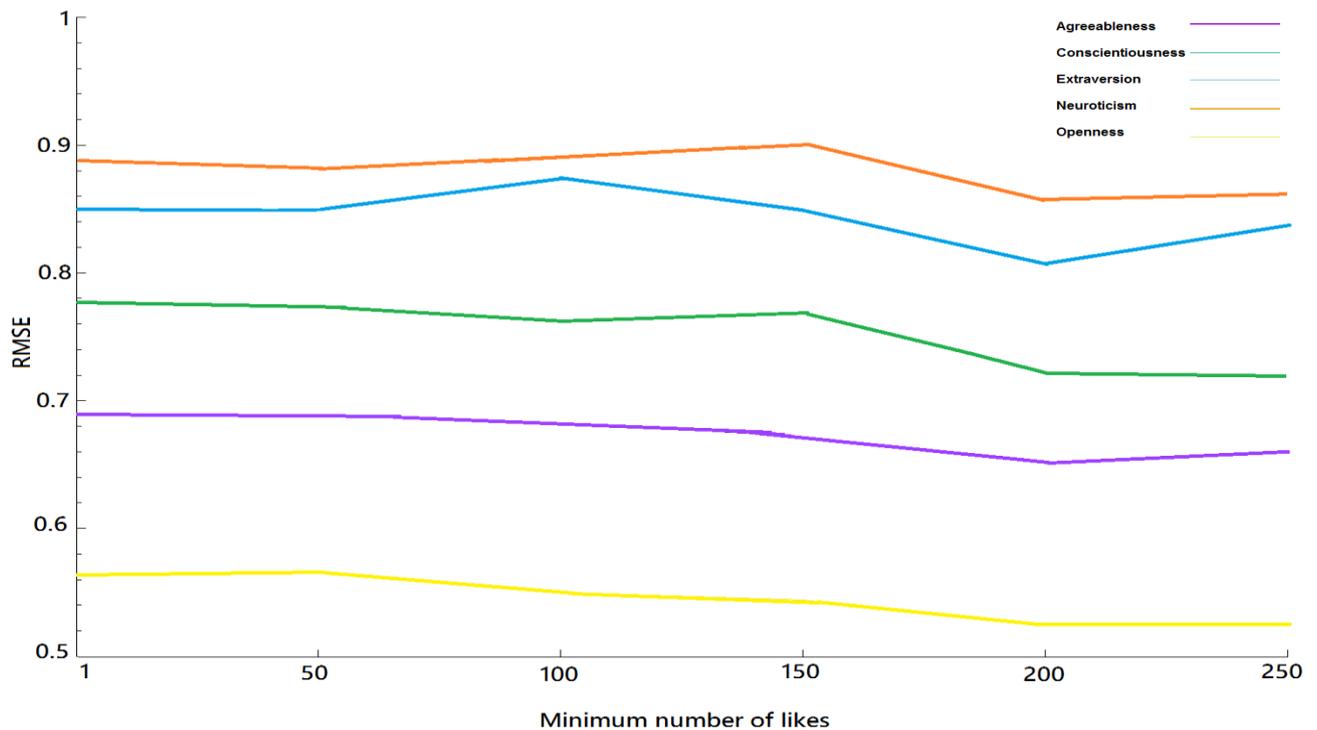

Figure 6. Root Mean Squared Error (RMSE) value for varying minimum number of likes with fixed-size training set.

## CONCLUSION

We reported the feasibility of modelling the Big 5 personality traits of users based on extracting metadata of pages users liked on Facebook. We used the hierarchy that Facebook employs to categorize pages as features to train our machine learning models to effectively predict the Big Five personality scores. This allows for a quite accurate assessment of a person's personality traits and can be used in a wide variety of applications. While the prediction performance differs between the traits, our results show that we can predict the personality scores within 8-15% of the actual value and we developed a web-based prototype to access this prediction. The insights we gained when doing our research can now be combined with the results of other researchers to create a learning ensemble that can predict the personality of a social media users to a very precise approximation.

---

[4] http://wiki.dbpedia.org/

# REFERENCES

1. Amichai–Hamburger, Y., Lamdan, N., Madiel, R., & Hayat, T. (2008). Personality characteristics of Wikipedia members. CyberPsychology & Behavior, 11(6), 679-681.

2. Back, M. D., Stopfer, J. M., Vazire, S., Gaddis, S., Schmukle, S. C., Egloff, B., & Gosling, S. D. (2010). Facebook profiles reflect actual personality, not self-idealization. Psychological science, 21(3), 372-374.

3. Bin Tareaf, R., Berger, P., Hennig, P., Jung, J., & Meinel, C. (2017). Identifying Audience Attributes: Predicting Age, Gender and Personality for Enhanced Article Writing. In Proceedings of the 2017 International Conference on Cloud and Big Data Computing (pp. 79-88). ACM.

4. Costa, P. T., & McCrae, R. R. (1989). NEO five-factor inventory (NEO-FFI). Odessa, FL: Psychological Assessment Resources.

5. Farnadi, G., Sitaraman, G., Rohani, M., Kosinski, M., Stillwell, D., Moens, M. F., & De Cock, M. (2014). How are you doing?: emotions and personality in Facebook. In EMPIRE2014 (2nd Workshop on" Emotions and Personality in Personalized Services"), workshop at UMAP2014 (22nd Conference on User Modelling, Adaptation and Personalization) (pp. 45-56).

6. Farnadi, G., Sitaraman, G., Sushmita, S., Celli, F., Kosinski, M., Stillwell, D., & De Cock, M. (2016). Computational personality recognition in social media. User modeling and user-adapted interaction, 26(2-3), 109-142.

7. Gosling, S. D., Augustine, A. A., Vazire, S., Holtzman, N., & Gaddis, S. (2011). Manifestations of personality in online social networks: Self-reported Facebook-related behaviors and observable profile information. Cyberpsychology, Behavior, and Social Networking, 14(9), 483-488.

8. Kosinski, M., Matz, S. C., Gosling, S. D., Popov, V., & Stillwell, D. (2015). Facebook as a research tool for the social sciences: Opportunities, challenges, ethical considerations, and practical guidelines. American Psychologist, 70(6), 543.

9. Goldberg, L. R. (1993). The structure of phenotypic personality traits. American psychologist, 48(1), 26.

10. Sewwandi, D., Perera, K., Sandaruwan, S., Lakchani, O., Nugaliyadde, A., & Thelijjagoda, S. (2017, January). Linguistic features based personality recognition using social media data. In Technology and Management (NCTM), National Conference on (pp. 63-68). IEEE.

11. Thilakaratne, M., Weerasinghe, R., & Perera, S. (2016, October). Knowledge-driven approach to predict personality traits by leveraging social media data. In Web Intelligence (WI), 2016 IEEE/WIC/ACM International Conference on (pp. 288-295). IEEE.

12. Zhao, S., Grasmuck, S., & Martin, J. (2008). Identity construction on Facebook: Digital empowerment in anchored relationships. Computers in human behavior, 24(5), 1816-1836.